# Designing Privacy for You
## A Practical Approach for User-Centric Privacy

Awanthika Senarath[1], Nalin A.G. Arachchilage[2], and Jill Slay[3]

[1] Australian Centre for Cyber Security
University of New South Wales - Canberra
Australian Defence force Academy, Australia
Email: a.senarath@student.unsw.edu.au
[2] Email: nalin.asanka@adfa.edu.au
[3] Email: jill.slay@adfa.edu.au

**Abstract.** Privacy directly concerns the user as the data owner (data-subject) and hence privacy in systems should be implemented in a manner which concerns the user (user-centered). There are many concepts and guidelines that support development of privacy and embedding privacy into systems. However, none of them approaches privacy in a user-centered manner. Through this research we propose a framework that would enable developers and designers to grasp privacy in a user-centered manner and implement it along with the software development life cycle.

## 1 Introduction

Donald Trump's presidential campaign was seriously damaged, when a personal conversation he had with a friend ten years ago (in 2005) was recorded and released to the public during his run in the elections in 2016 in the USA. The recording had been done without the knowledge of the two people engaged in the conversation and was released at a very critical moment in the presidential campaign [1]. This is not much different to the situation users face on a daily basis when using on-line applications to network, communicate, shopping and banking on-line, and for many other personal tasks [9]. Due to the pervasiveness of Information and Communication Technology on-line applications have become an integral part of users [7]. However, they are unaware of the plethora of sensitive information that is being collected by applications in the background, the entities that have access to those information and how securely their data is stored [40], because those systems are not designed for end user privacy. Therefore, users unknowingly become vulnerable to highly personalized scam, and identity theft [41]. For example, a cyber criminal called Peace listed 200 million of user records of Yahoo users for sale on the dark web at the beginning of August, 2016, which consisted of user names, encrypted passwords [3].

Caputo et al [10] in their investigation on barriers to usable systems, claim that companies do attempt to adhere to theories that improve usability of systems to secure company reputation for protecting market shares. Obstructing existing user practices in systems is mentioned as one of the five pitfalls by Lederer et al [31] that should be avoided by privacy designers to ensure usable privacy. However, other than providing lengthy, in-comprehensive data policies and user guides [44], [2], very little attention has been paid by organizations to integrate privacy as a part of user engagement with the system. A usable privacy implementation would help users better understand the system and manage their personal boundaries in interacting with systems. However, up-to-date security and privacy is a different process in an application, which the user is forced to go through [44], [40].

Privacy directly concerns the user as the data owner [8]. Privacy of a system should be designed for the user (user-centered) [5] and to be *user-centric*, designers should take a step further to analyze users' expected behavioral engagement with the system, to ensure they address potential privacy risks [49]. However, current approaches for privacy design mostly concern data as an impersonalized entity [22] and ignores users perspectives and behavioral traits in embedding privacy into systems [52]. For example, Caputo et al [10] has pointed that developers have different perceptions on what usability really means, and also thinks "developers know best", and see little or no need to engage with target users. As a solution to this problem, Rubenstien et al [42] highlights the importance of extending the existing user experience (UX) evaluation techniques to improve usability of privacy implementations in systems. We are contributing by providing a systematic approach for software developers to design privacy as they approach the software development lifecycle with a user-centered view [48], [17]. We propose a paradigm shift in developer thinking from *Implementing Privacy in a System* to *Implementing Privacy for Users of the System*, through the concept of user-centered privacy.

## 2    Related Work

There are many conceptual and technological guidelines introduced by researchers to support developers and designers to implement privacy in applications. Fair information practices (FIP) [47], Privacy by Design (PbD) [11], Privacy Enhancing Tools (PET) [24] and Privacy Impact Assessment (PIA) [14] are guidelines or principles that have emerged to support developers to implement privacy in systems. However, we are experiencing major privacy failures in systems and applications [29] because, these guidelines are not formed in a way that is practically applicable with the software development processes today [17]. There is a gap in privacy guidelines for developers and privacy in practice [50].

Fair Information Practices (FIP) focus on the rights of individuals, and the obligations of institutions associated with the transfer and use of personal data such as data quality, data retention and notice and consent/choice of users [47]. Here, personal data is data that are directly related to the identification of a person and their behaviors [47]. FIP is criticized to lack comprehensiveness in scope and to be unworkable and expensive [19]. This is where PbD gained its recognition. It is fair to say that PbD is an improved state of FIP [17] with focus on a wider range of requirements considering the business goals of the company [11]. PbD was introduced as a set of guidelines to embed privacy as a part of system design in the designing phase it-self [15]. It involves seven principles [11] which focus on the developer and the company perspective of privacy rather than user perspective [22], [52], [11]. For example, the last principle in PbD states that it should respect for user privacy, however it does not tell how to design privacy in a user-centric way [22], [48]. Furthermore, due to the widespread of the PbD concept, it has become a fashionable idea to declare commitment to PbD, without genuine privacy interests [17]. Therefore, Davies et al [17] highlights the importance for an integrated approach to developing PbD as a practical framework to overcome its flaws, as otherwise PbD would remain the accepted norm comprehended only by a small community of good privacy practitioners.

Privacy Impact Assessment (PIA) is a more practical approach that focus on the impact of privacy on the stakeholders of a project [14]. However, PIA is only a first step and should be followed by PbD and implementation technologies for completeness. Privacy Enhancing Technologies (PETs) are used to implement certain designs to overcome the risks identified in the PIA [24]. However, PETS can be used only after careful analysis of privacy requirements and following a systematic approach to decide how privacy should be implemented in the system. That is where the framework we propose is going to fit in. While PETs are useful in gaining insights into the technological capability in implementing privacy in systems, PbD provides the base on which privacy should be designed prior to implementation, bridging the former with latter is a timely requirement [17].

PRIPARE [30], a recent development in the field of privacy engineering, elaborates how PbD should be implemented in practice. This work address the part that was lacking in PbD in great detail. They consider a similar approach to ours defining how PbD should be applied in each step of a generic development cycle, considering environmental factors to assist development throughout. However, PRIPARE, too considers privacy only from the perspective of the application developers and organizations. Even though they encourage respect for user privacy similar to PbD, they fail to nudge developers to think in a user-centered manner, comprehend real user requirements in privacy, varying on the nature and purpose of the application, context of usage of the application, and also on the characteristics of the user group. We propose *User-Centered Privacy* with all steps in the framework proposed, centered on the user of the system.

In the framework for "Engineering privacy" [49], which proposes a solution for practical privacy implementations, the importance of understanding user behaviors in implementing privacy is heavily discussed. However, as they have proposed in their framework, it is not realistic to implement privacy-by-architecture in a system where annoymity, data-minimization and other privacy technologies are practiced to an extent where providing information, notice, and choice and access to users can be completely ignored [24]. On the other hand, Cannon [9] has provided a comprehensive guide for implementing privacy in a system addressing all parties in a company. It involves a very descriptive data analysis framework and a guide for privacy related documentation. However, this lacks a user-centric approach towards privacy and focuses solely on the development perspective. It considers data from the company perspective and ignores user's behavioral engagement with the system and their expectations and perceptions of privacy. Furthermore, it is not defined adhering to a particular development process, or a work-flow and only contains best practices that should be followed by developers for embedding privacy into systems. Therefore it is not possible for an organization to directly apply them to the current development processes they practice. To address both these gaps we have implemented our framework on Unified Software Development Process (UP), created by Jacobson, Booch, and Rambough [18].

Iterative and Incremental software development processes ( [4], [45], [23]) are highly used today in organizations [13] due to their capability of handling varying requirements in short time periods. UP is the most descriptive form of Iterative and Incremental development processes from which the modern light scale development processes customized for has been derived from [43]. It defines steps not only for the development of a software application, but also to manage, maintain and support it throughout [43]. Therefore, we define the proposed framework (Figure 1) on UP so that it could be easily linked to the lightweight simplified interactive and incremental software development processes that are used widely.

## 3   Systematic Approach for User-Centric Privacy

The proposed framework considers the phases defined in UP through which the project moves over time, with each phase containing balanced out amounts of analyzing, implementation and designing tasks involved. The four phases in the UP life cycle are inception, elaboration, construction and transition [43]. The proposed framework defines tasks to be carried out in each of these phases, so that privacy would be a part of the development process throughout.

For example, consider developing a mobile gaming application. A privacy risk estimation for all stakeholders such as game players, developers and the company that releases the game as well as the platform that hosts the game should be done in the inception phase. Afterwards, effective data minimization in the inception phase would ensure commitment to privacy and better understanding of privacy requirements from beginning itself. This is expected in PbD, as integrating privacy at later phases would not deliver expected results in terms of privacy [11]. Analyzing of data gathering requirements and players behaviors, and the environment in which the game would be played (on phone, tablet in public places), setting privacy goals and high privacy risk mitigation designs in the elaboration phase would further strengthen the company's privacy goals and ensure usability of privacy designs [49]. Identifying any remaining privacy requirements, reviewing, user surveys and comparing players' expectations against implementation for preparing privacy policies is required in the construction phase. This would aid effective transparency in the game application being designed [25]. Testing privacy, defining privacy setting for deployment guide and accountability evaluation [12] at the transition phase would sum up the work-flow for user-centric privacy implementation in the game.

Figure 1 describes the proposed framework as a work-flow, that should be followed by the development team collectively, to achieve privacy in the system. Each step in the work-flow is tightly bound to the next, such that the results and knowledge gained in the initial step assist the execution of the next step. Environment and change management are specifically defined in UP to support planning and management of the project [43].This is an essential step in software development given the continuous change requirements and modification that happens in practice. We have hence included these steps in our work-flow to ensure continuous privacy commitment.

We expect to conduct a study involving application developers and end users to validate our framework, and to receive feedback to fine tune it to improve its potential for practical realization. Sections of the questionnaire that aligns with each step are embedded to show how we aim to validate the steps proposed. The full questionnaire of the study is available in the appendix.

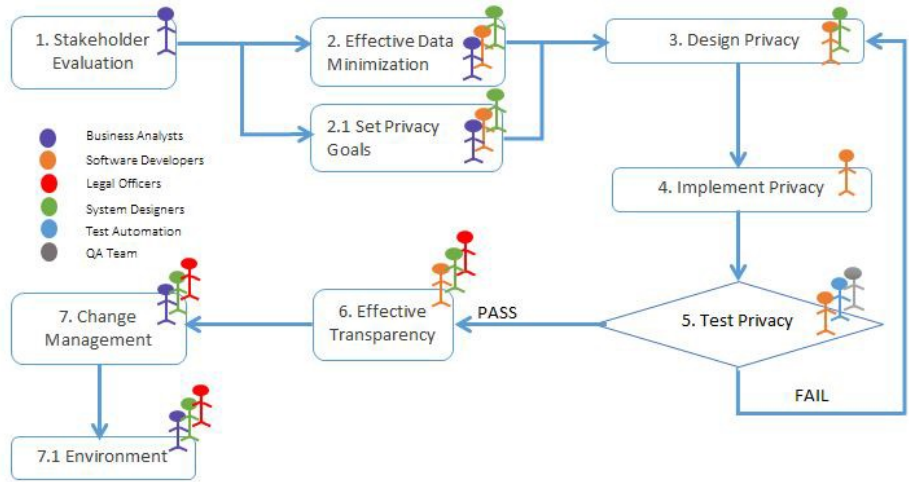

**Fig. 1.** User-Centric Privacy Framework

**1. Stakeholder Evaluation :** PIA [24] already proposes a privacy impact analysis on the end user in designing a system. However, we propose to assess both the company and the users rather than just assessing the impact of the system to the end user. For effective privacy, all stakeholders of the system need to be analyzed in terms of their privacy expectations, responsibility and potential vulnerabilities [14], to understand their requirements, perceptions and behaviors with the system. Users and the company as an entity should be considered for their expected goals from the system being developed, their expected engagement with the system and the potential privacy impact that could arise and accountability [12]. Accountability is considered to be a strong parameter in effective privacy implementation as it ensures reliable and responsible systems [12]. A stakeholder evaluation report should be generated at the end of the evaluation and the report should be composed to be available during the latter stages. Understanding the stakeholders in terms of privacy would help designers for effective data minimization with a better view of the users expectations.

**2. *User-Centric* Data Taxonomy for Privacy :** *Data minimization* is a very broad statement in FIP [8]. However for effective data minimization usage of Data should be minimized in a meaningful way. Collecting a small amount of highly sensitive data that is irrelevant to the purpose of the application cannot be voted as good compared to collecting a large amount of less sensitive data related to the application. For example users would expect a health care website to store their past health conditions, but not for a social networking site [40]. It was shown that users were comfortable with web-sites collecting their data when it directly relates to the purpose they serve [34]. To this end we are proposing the *Data Taxonomy for Privacy* for effective data minimization.

Barker et al [8] proposes a data taxonomy for privacy which considers three dimensions for data namely, purpose, visibility and granularity. Based on the same concept we propose a taxonomy with purpose (relevance) and sensitivity and visibility. We believe that for a user-centric approach, sensitivity of data elements and their visibility in the application are important parameters [37]. *Sensitivity* could be defined as the risk involved in exposing a particular data element to public, and *visibility* is the exposure that data element has by default in the application [36]. Cannon's data analysis framework classifies data that is being collected depending on their exposure, consent and user awareness. However, it lacks analyzing and differentiating data categories with the scope and purpose of the application and the sensitivity of the data elements [9]. Our data taxonomy is created learning from these classifications. Following are the steps to follow for effective data minimization.

- Step 1 : Categorize the application according to their purpose. The application could be performing *social networking (Facebook), communication (Online Chatting, Video Calling), gaming and entertainment, health or household (Home IoT).*
- Step 2 : Depending on the category of the application rank the users they expect to collect in the order of relevance to the purpose of the application [14].
- Step 3 : Categorizing user-data you expect to collect from user perspective, such as *Personal Identification Data (Name, Age!), Behavioral Data collected from user (purchasing, health behavior), Financial Data, Societal Data (Occupation, Marital/Family Status) and History (Travel, Health, Occupation).*
- Step 4 : Score all data according to their sensitivity and visibility and rank : In rating privacy in Facebook, Minkus et al [37] has shown that the privacy level of certain types of data depends on its sensitivity to the data subject and its visibility in the given application [36]. The framework already put forward by Liu et al [35] to evaluate sensitivity and visibility of data elements can be applied here.

- Step 5 : Looking at the rankings, if the application is collecting a data type that is less relevant to the purpose and scope of their application, which are not directly required to achieve their business goals, and have a higher sensitivity ranking, either take measures to improve their data collection strategy or improve their application to ensure access, choice and notice for those highly sensitive data elements [9].

Through effective data minimization, analysts should aim to ensure a win-win situation in achieving both privacy and business goals as defined in PbD [14] in the designing step of the development life-cycle.

---

Validation Questions Proposed:

- a. What are the data elements that you collect in this application?
- b. Rank the data elements you collect in the order of relatedness, sensitivity and visibility of those data elements to the purpose the application serves
- c. For those highly sensitive data elements that are less relevant to the purpose of the application,
    - i. Is there any way that you can avoid collecting those data?
    - ii. What is the database structure they being stored? Have you evaluated the privacy risk factor in designing these databases?
    - iii. Have you considered hiding, separation, aggregation, notice and control in designing the databases?
    - iv. What are the risk mitigation strategies that you have specifically used?

Consider yourself as an end-user of the application you designed,

- a. What data do you expect this application to collect?
- b. What data are you willing to expose to the application for your benefit as a user? Create separate lists imagining you as a health-care professional registered to provide service and a user seeking medical advice.
- c. What sort of data do you think the application should specifically refrain from storing and making use of? Why?
- d. What data do you believe the application collected in the background while you were using the application?

---

**3. Set Privacy Goals and Design Privacy :** In designing privacy goals for their system designers should take into consideration how a user is expected to engage with the application they design. The amount of data the user is going to expose and their expected level of privacy are highly dependent on users' engagement with the application [38], [20]. Also the usability and adaptability of the privacy enhancing tools designers embed in the system largely depends on users behavioral engagement of the system. If the designers place the privacy tools in an accessible way, but not visible to the user in their natural engagement with the application it is not likely to be used. Hence we emphasize that designers and developers should focus on the behavioral engagement of users with the system, similar to how User-experience (UX) designers test and evaluate their interfaces [51]. In terms of privacy goals the designers should separately consider users' privacy goals and the company's business goals. User goals could be defined through a user-survey. As defined in PbD concepts, it is important to see privacy and business goals as common goals which should not be compromised for each other [11]. As proposed in our *effective data minimization* above and *transparency with privacy policies*, which follows the designing steps, we have shown how to achieve this win-win situation in practice.

Designing privacy is the essence of PbD concept [11]. This involves on defining the data access, retaining policies and storage of data. Through *data minimization work-flow*, designers would get an idea of the sensitivity and relativity of the data elements they access, which gives them a better position to effectively decide on the consent/choice/access options they should embed in the system. For this it is important that they understand the user behavior and user perception of privacy. Spiekermann et al [49] in their framework emphasize the importance of developers understanding the *user sphere* in implementing effective privacy. User sphere means the user perception, how privacy could be breached, how users see and behave with the system. Similarly in our user-centric privacy designing approach we stress the importance of developers understanding potential vulnerabilities for the user, and the company and data breaches from user perspective and company perspective. Developers could understand these aspects through user-surveys and interviews. Defining user-centric privacy goals and designs would support developers to implement privacy into the system in a user-centric manner.

**4. Implementing Privacy :** Developers can incorporate existing PETs wherever applicable in achieving the privacy goals set forth by the designers. There are ample PETs that has been designed so far and technologies that are adopted to implement privacy in systems [16]. These includes mechanisms for anonymity, network invasion, identity management, censorship resistance, selective disclosure credentials and also database implementation to preserve privacy [16]. Selection of PETs are highly subjective and dependent on the privacy goals and requirements of the software being developed [24]. It is beyond the scope of this paper to explicitly discuss existing PETs. However, a fact worth noticing is that it is not possible to achieve a 100% privacy preserved system through pure architectural and technological support as suggested in the Engineering privacy framework by Spiekermann [49]. Privacy in a system should be achieved through a balanced approach with privacy architecture, policy implementation, communication and transparency as guided in our work-flow for user-centric privacy. All the implementations and designs should be tested in a process that involves real end users and other entities that are devoted for testing applications as explained in the following section.

**5. Testing for Privacy :** Testing is the most important section in software development. In the proposed framework we define the following guidelines to be followed in testing for privacy with a *user-centric* approach. Testing should follow the privacy implementation steps. A cycle of designing and implementation should follow in terms of failure of any of the following guidelines.

– Preserving privacy of data in applications during testing in database-centric applications (DCAs) : It is argued that data anonymization algorithms as k-anonymity [32], taken to preserve database privacy seriously degrades testability of applications [21]. However, guessing-anonymity [39], a selective anonymity metric, is voted better against other forms due to the possibility of selective anonymization.

– Testing the application for potential privacy vulnerabilities : The privacy risk assessment and the information flow diagrams could be used by testers to gain an idea of potential privacy vulnerabilities. Information collections, processing, dissemination and invasion are identified by Solove as the four stages in an application where privacy could be breached [46]. QA teams and test automation teams should test the application for potential privacy vulnerabilities in these stages.

– Testing the usability of implemented privacy specific tools (privacy user-setting processes) : During the initial phases of development there could be tools incorporated in the system, explicitly to reduce privacy risks. Usability of these tools should be tested against real users during the test phase to ensure their effectiveness.

**6. *User-centric* approach for Transparency :** In the current context transparency means displaying users about what data is being collected, how the data is going to be retained and used in the privacy policy [26]. However, almost none of the applications today is successful in effectively displaying these data through the privacy policy [29]. The privacy policy is incorporated by many companies as a tool to comply to legal obligations [26]. However, we believe that in transparency the true meaning is not just displaying what the application does, but also to bridge the gap between user expectation versus reality [25]. Companies can win the trust of users and users would be more open and comfortable using applications as they have knowledge on what is happening to their data in the application [33]. For this developers should perform a survey and a usability study of the developed system prior to defining their privacy policies. Rao et al, [40] in their study explains the mismatched user expectations on-line. Based on this we propose the user-centric approach for transparency through evaluation of user expectations versus real design. This way companies can get more accurate details about users, and win users trust while achieving their business goals [34], which is the one of the principles in PbD [11]. The proposed work-flow is,

– Step 1 : List collection/retain and storage of data in the application.
– Step 2 : Conduct a user survey with the application that covers general use-cases; information about the data the user expects the system to collect, users' understanding on how the data is stored/used by the system.
– Step 3 : Identify the mismatches between user expectation versus reality; generate privacy policy with details covering discovered mismatches.
– Step 4 : Conduct a readability test and evaluate the privacy policy, Flesch readability test could be used for this purpose [28] (A model that is widely used text readability)
– Step 5 : Publish the privacy policy, minimize changes. In the case of unavoidable changes inform users well in advance.

---

Validation Questions Proposed:

– a. What information you believe the users would expect to see in the privacy policy and why?
– b. What information have you included in the privacy policy specifically to oblige for legal requirements?
– c. What information have you included in the privacy policy specifically to maintain transparency and better communication with the end user?
– d. What is the readability score of the privacy policy? (Use the Felsch Readability evaluation formula)
– e. Do you think navigation in the privacy policy an important feature?

Consider yourself as an end user of the application you design.

– a. Do you think this application should have a privacy policy? Why?
    – i. Would you take time to go through the privacy policy of the application?
– b. Write down things you wish to see in the privacy policy of this application.
– c. Write down the most important sections you believe that should be covered in the privacy policy of this application.

---

**7. Documentation :** Documentation should be approached with the focus of *What information do developers need from past projects?* [53]. Cannon et al [9] has specified documentation requirements for privacy maintenance. However, in UP the focus is more on creating and maintaining models rather than textual documentation for quick adaptability and change management [53]. The benefit of the documents should always overweight the cost of creating it and models are encouraged as temporary documents that are discarded once their purpose is served [6]. Based on Cannon's suggestions considering the current context, we propose generating the following documents with relation to privacy design,

– Stakeholder privacy risk evaluation report : during inception
– Data flow diagram : during inception
– Privacy statement for the application end user : during transition
– Deployment guide with privacy settings : during construction
– Review document about privacy issues, risk mitigation and responsible parties in decisions made (Accountability) : during transition

PbD emphasizes the importance of adopting privacy at the earliest stage in system design [11]. We highlight the importance of adopting privacy design early as well as continuing it until the very last step of the software development life-cycle. We show clearly how to achieve that in practice with a user-centered approach through a comprehensive step by step guide. Software development is rarely a single process that involves a single person [27]. Our privacy framework as shown in image 1 comprehensively captured the role of each party in terms of privacy design and implementation [27]. Most importantly the proposed framework, coins the term *User-Centered Privacy* and comprehensively emphasize how developers should adopt a user centered mentality in approaching privacy in systems.

## 4   Conclusion and Future Work

As on-line applications are being used to achieve simple day to day tasks, using them is not something users could refrain from due to privacy concerns. As users are getting more and more concerned about their privacy on-line, developers should focus on embedding privacy right into their applications with a user-centric approach. Through this paper, we contribute *A practical workflow to implement user-centric privacy design*, which is a timely requirement for effective privacy designing and implementation [17]. To the best of the authors' knowledge, this is the first framework designed to specify an end-to-end work-flow to achieve privacy as a user-centric approach with current software development processes.

Interviewing software developers and designers to understand their expectations and understandings on implementing privacy is highly desirable for strengthening and fine tuning the proposed framework in a more pragmatic manner. Applying the framework to more abstract and practical development processes like agile, scrum would also help fine tuning.

## References

1. David A. Fahrentholdtrump recorded having extremely lewd conversation about women in 2005. https://www.washingtonpost.com/politics/trump-recorded-having-extremely-lewd-conversation-about-women-in-2005/2016/10/07/3b9ce776-8cb4-11e6-bf8a-3d26847eeed4_story.html. Accessed: 2016-10-14.
2. Facebook data policy. https://www.facebook.com/policy.php. Accessed: 2016-08-21.

3. Paul Curran august 2016 hacks: 8 of the largest hacks, breaches and cyber incidents. https://www.checkmarx.com/2016/09/11/august-2016-hacks-8-largest-hacks-breaches-cyber-incidents. Accessed: 2016-11-03.

4. P. Abrahamsson, O. Salo, J. Ronkainen, and J. Warsta. Agile software development methods: Review and analysis, 2002.

5. A. Adams and M. A. Sasse. Users are not the enemy. *Communications of the ACM*, 42(12):40 – 46, 1999.

6. A. Anwar. A review of rup (rational unified process). *International Journal of Software Engineering (IJSE)*, 5(2):12 – 19, 2014.

7. N. A. G. Arachchilage and A. P. Martin. A trust domains taxonomy for securely sharing information: A preliminary investigation. In *HAISA*, pages 53 – 68, 2014.

8. K. Barker, M. Askari, M. Banerjee, K. Ghazinour, B. Mackas, M. Majedi, S. Pun, and A. Williams. A data privacy taxonomy. In *British National Conference on Databases*, pages 42 – 54. Springer, 2009.

9. J. Cannon. *Privacy: what developers and IT professionals should know*. Addison-Wesley Professional, 2004.

10. D. D. Caputo, S. L. Pfleeger, M. A. Sasse, P. Ammann, J. Offutt, and L. Deng. Barriers to usable security? three organizational case studies. *IEEE Security & Privacy*, 14(5):22 – 32, 2016.

11. A. Cavoukian. Privacy by design: the definitive workshop. a foreword by ann cavoukian, ph. d. *Identity in the Information Society*, 3(2):247 – 251, 2010.

12. A. Cavoukian, S. Taylor, and M. E. Abrams. Privacy by design: essential for organizational accountability and strong business practices. *Identity in the Information Society*, 3(2):405 – 413, 2010.

13. J. Cho. A hybrid software development method for large-scale projects: rational unified process with scrum. *Issues in Information Systems*, 10(2):340 – 348, 2009.

14. R. Clarke. Privacy impact assessment: Its origins and development. *Computer law & security review*, 25(2):123 – 135, 2009.

15. G. Danezis, J. Domingo-Ferrer, M. Hansen, J.-H. Hoepman, D. L. Metayer, R. Tirtea, and S. Schiffner. Privacy and data protection by design-from policy to engineering. *arXiv preprint arXiv:1501.03726*, 2015.

16. G. Danezis and S. Gürses. A critical review of 10 years of privacy technology. *Proceedings of surveillance cultures: a global surveillance society*, pages 1 – 16, 2010.

17. S. Davies. Why privacy by design is the next crucial step for privacy protection, 2010.

18. A. Fuggetta. Software process: a roadmap. In *Proceedings of the Conference on the Future of Software Engineering*, pages 25 – 34. ACM, 2000.

19. R. Gellman. Fair information practices: A basic history. *Available at SSRN 2415020*, 2015.

20. M. N. Giannakos, K. Chorianopoulos, K. Giotopoulos, and P. Vlamos. Using facebook out of habit. *Behaviour & Information Technology*, 32(6):594 – 602, 2013.

21. M. Grechanik, C. Csallner, C. Fu, and Q. Xie. Is data privacy always good for software testing? In *2010 IEEE 21st International Symposium on Software Reliability Engineering*, pages 368 – 377. IEEE, 2010.

22. S. Gürses, C. Troncoso, and C. Diaz. Engineering privacy by design. *Computers, Privacy & Data Protection*, 14(3), 2011.

23. J. G. Hall and L. Rapanotti. Towards a design-theoretic characterisation of software development process models. In *Proceedings of the Fourth SEMAT Workshop on General Theory of Software Engineering*, pages 3 – 14. IEEE Press, 2015.

24. J.-H. Hoepman. Privacy design strategies. In *IFIP International Information Security Conference*, pages 446 – 459. Springer, 2014.

25. C. Jensen and C. Potts. Privacy policies as decision-making tools: an evaluation of online privacy notices. In *Proceedings of the SIGCHI conference on Human Factors in Computing Systems*, pages 471 – 478. ACM, 2004.

26. P. G. Kelley, L. Cesca, J. Bresee, and L. F. Cranor. Standardizing privacy notices: an online study of the nutrition label approach. In *Proceedings of the SIGCHI Conference on Human factors in Computing Systems*, pages 1573 – 1582. ACM, 2010.

27. J. Y. T. Kerry Spalding. Practical strategies for integrating privacy by design throughout product development process. In *Proceedings of the 2016 CHI Conference Extended Abstracts on Human Factors in Computing Systems*, pages 3415 – 3422. ACM, 2016.

28. J. P. Kincaid, R. P. Fishburne Jr, R. L. Rogers, and B. S. Chissom. Derivation of new readability formulas (automated readability index, fog count and flesch reading ease formula) for navy enlisted personnel. Technical report, DTIC Document, 1975.

29. P. Kumar. Ranking digital rights: Pushing ict companies to respect users privacy. In *CHI 16 Workshop: Bridging the Gap between Privacy by Design and Privacy in Practice*, page 15. CHI, 2016.

30. A. Kung, A. C. Garcia, N. N. McDonnell, I. Kroener, D. Le Métayer, C. Troncoso, and Y. S. M. UPM. Pripare: A new vision on engineering privacy and security by design. 1, 2016.

31. S. Lederer, J. I. Hong, A. K. Dey, and J. A. Landay. Personal privacy through understanding and action: five pitfalls for designers. *Personal and Ubiquitous Computing*, 8(6):440 – 454, 2004.

32. K. LeFevre, D. J. DeWitt, and R. Ramakrishnan. Incognito: Efficient full-domain k-anonymity. In *Proceedings of the 2005 ACM SIGMOD international conference on Management of data*, pages 49 – 60. ACM, 2005.

33. P. G. Leon, A. Rao, F. Schaub, A. Marsh, L. F. Cranor, and N. Sadeh. Why people are (un) willing to share information with online advertisers, 2015.

34. P. G. Leon, B. Ur, Y. Wang, M. Sleeper, R. Balebako, R. Shay, L. Bauer, M. Christodorescu, and L. F. Cranor. What matters to users?: factors that affect users' willingness to share information with online advertisers. In *Proceedings of the ninth symposium on usable privacy and security*, page 7. ACM, 2013.

35. K. Liu and E. Terzi. A framework for computing the privacy scores of users in online social networks. *ACM Transactions on Knowledge Discovery from Data (TKDD)*, 5(1):6, 2010.

36. E. M. Maximilien, T. Grandison, T. Sun, D. Richardson, S. Guo, and K. Liu. Privacy-as-a-service: Models, algorithms, and results on the facebook platform. In *Proceedings of Web*, volume 2, 2009.

37. T. Minkus and N. Memon. On a scale from 1 to 10, how private are you? scoring facebook privacy settings. In *Proceedings of the Workshop on Usable Security (USEC 2014). Internet Society*, 2014.

38. A. Oulasvirta, T. Rattenbury, L. Ma, and E. Raita. Habits make smartphone use more pervasive. *Personal and Ubiquitous Computing*, 16(1):105 – 114, 2012.

39. Y. Rachlin, K. Probst, and R. Ghani. Maximizing privacy under data distortion constraints in noise perturbation methods. In *Privacy, security, and trust in KDD*, pages 92 – 110. Springer, 2009.

40. A. Rao, F. Schaub, N. Sadeh, A. Acquisti, and R. Kang. Expecting the unexpected: Understanding mismatched privacy expectations online. Federal Trade Commission PrivacyCon Conference, 2016.

41. H. Rashtian, Y. Boshmaf, P. Jaferian, and K. Beznosov. To befriend or not? a model of friend request acceptance on facebook. In *Symposium On Usable Privacy and Security (SOUPS 2014)*, pages 285 – 300, 2014.

42. I. Rubinstein and N. Good. Privacy by design: A counterfactual analysis of google and facebook privacy incidents. 2012.

43. J. W. Satzinger, R. B. Jackson, and S. D. Burd. *Object-oriented Analysis and Design: With the Unified Process*. Thomson Course Technology, 2005.

44. F. Schaub, R. Balebako, A. L. Durity, and L. F. Cranor. A design space for effective privacy notices. In *Eleventh Symposium On Usable Privacy and Security (SOUPS 2015)*, pages 1 – 17, 2015.

45. K. Schwaber. Scrum development process. In *Business Object Design and Implementation*, pages 117 – 134. Springer, 1997.

46. D. J. Solove. A taxonomy of privacy. *University of Pennsylvania law review*, pages 477 – 564, 2006.

47. D. J. Solove, M. Rotenberg, and P. M. Schwartz. *Information privacy law*. 2010, 4th edition.

48. S. Spiekermann. The challenges of privacy by design. *Communications of the ACM*, 55(7):38 – 40, 2012.

49. S. Spiekermann and L. F. Cranor. Engineering privacy. *IEEE Transactions on software engineering*, 35(1):67 – 82, 2009.

50. L. Stark, J. King, X. Page, A. Lampinen, J. Vitak, P. Wisniewski, T. Whalen, and N. Good. Bridging the gap between privacy by design and privacy in practice. In *Proceedings of the 2016 CHI Conference Extended Abstracts on Human Factors in Computing Systems*, pages 3415 – 3422. ACM, 2016.

51. R. Unger and C. Chandler. *A Project Guide to UX Design: For user experience designers in the field or in the making*. New Riders, 2012.

52. J. van Rest, D. Boonstra, M. Everts, M. van Rijn, and R. van Paassen. Designing privacy-by-design. In *Annual Privacy Forum*, pages 55 – 72. Springer, 2012.

53. S. Voigt, J. von Garrel, J. Müller, and D. Wirth. A study of documentation in agile software projects. In *Proceedings of the 10th ACM/IEEE International Symposium on Empirical Software Engineering and Measurement*, page 4. ACM, 2016.